\documentclass[twocolumn, floatfix, showkeys, showpacs, article, asp]{revtex4-1}
\usepackage[utf8]{inputenc}
\usepackage{amsmath}
\usepackage{amsfonts}
\usepackage{amssymb}
\usepackage{graphicx}
\usepackage{textcomp}
\usepackage{subfigure}
\usepackage[margin=.6in]{geometry}
\usepackage{color}
\usepackage{float}
\usepackage{atbegshi,picture}
\usepackage{soul}
\usepackage{tikz}

\usepackage[pdfpagemode={UseOutlines}, bookmarks=true, bookmarksopen=true, bookmarksopenlevel=0, bookmarksnumbered=true, hypertexnames=false, colorlinks, linkcolor={blue}, citecolor={blue}, urlcolor={blue}, pdfstartview={FitV}, unicode, breaklinks=true]{hyperref}
\usepackage{multirow}
\begin{document}
\title{Spin-resolved Mott crossover and entanglement in the half-filled Hubbard model}
\author{Md Fahad Equbal}
\email{md179654@st.jmi.ac.in}
\author{M. A. H. Ahsan}
\email{mahsan@jmi.ac.in}
\affiliation{Department of Physics, Jamia Millia Islamia (Central University), New Delhi $110025$, India}
\date{\today}
\begin{abstract}
We investigate the interaction-driven reorganization of spin and charge correlations in finite Hubbard clusters using exact diagonalization. Focusing on half-filled and lightly doped square lattices, we analyze spin-resolved charge-gaps, local observables, two-point correlation functions, entanglement measures, principal component analysis (PCA) of correlation matrices and quantum-geometry-based distance metrics. At half-filling, we observe the emergence of a robust Mott gap whose spin-dependent component is controlled by the effective exchange energy scale $J\sim 4t^2/U$ at strong coupling, confirming that residual spin dynamics govern the separation between the lowest-spin and the next higher-spin charge excitation channels. Distinct cluster geometries and boundary conditions reveal how spin-singlet versus finite-spin ground-state influence charge and spin responses. Upon one-hole doping, the spin-resolved charge-gaps collapses, indicating restored compressibility and metallic behavior. PCA and quantum-geometry-based analysis provide complementary data-driven and wavefunction-based perspectives on correlation-driven Mott crossover phenomena. Our results demonstrate that finite-size Hubbard clusters exhibit clear signatures of the Mott physics, spin-charge interplay and emergent exchange energy scales, offering a unified microscopic picture of interaction-induced electronic reorganization. 

\end{abstract} 
\maketitle

\section{Introduction} 
\label{Intro}
The interaction-driven crossover from an itinerant metal to Mott insulator \cite{Mott1974, Giamarchi1997, Imada1998, Mukul2024}, remains one of the most fundamental and actively studied phenomena in strongly correlated electron systems. The Hubbard model \cite{Hubbard1963, Gutzwiller1963, Kanamori1963} serves as a minimal theoretical framework for capturing the essential physics of strong electronic correlations. At half-filling (one electron per site on average) on the square lattice, increasing on-site repulsion suppresses charge fluctuations and promotes Heisenberg-type local moment formation \cite{Anderson1959, Chao1977, Spalek2007}, leading to Mott insulating state with strong short-range antiferromagnetic correlations \cite{Imada1998, Rohringer2016, Maria2024}. In the thermodynamic limit, this crossover and its finite-temperature continuation have been extensively characterized using dynamical mean-field theory (DMFT) and its cluster extensions \cite{Katnelson2000, Kotliar2001, Park2008, Paki2019, Danilov2022, Downey2023}, determinant quantum Monte Carlo \cite{Thomas2021}, and diagrammatic extensions \cite{Riccardo2017, Kim2020}. These studies have established a detailed understanding of short-range correlations and collective ordering tendencies in the two-dimensional (2D) Hubbard model.

Despite this progress, finite-size Hubbard clusters remain of fundamental interest for two reasons. First, recent quantum simulation platforms and engineered nanostructures naturally realize small correlated clusters where finite-size and boundary effects are intrinsic rather than parasitic \cite{Inaba2013, Sotnikov2014, Christian2016, WuF2025}. Second, exact diagonalization (ED) provides full access to eigenstates and correlation functions \cite{Botzung2024, Fahad2025}, enabling the development and validation of data-driven and wavefunction-based diagnostics that are difficult to apply in large-scale many-body approaches. However, finite clusters also exhibit strong geometry- and spin-sector dependence \cite{Tokura2000}, raising the question of how generic correlation-driven crossovers remain when translational invariance and thermodynamic averaging are absent.

Recent work has emphasized that the Mott crossover is most sharply reflected not only in conventional observables, but also in nonlocal correlation patterns and fluctuation spectra \cite{Evgeny2021, Stepanov2022, Estepanov2022, Vandeli2024, Eastepanov2024, Lechermann2024}.
In particular, cluster DMFT investigations have demonstrated that short-range spin and charge correlations follow the Widom-line crossover \cite{Downey2023}, while real-space correlation studies have revealed the buildup of local and nonlocal correlations near the metal-insulator transition \cite{Maria2024}. These results motivate a complementary finite-cluster perspective where correlation matrices themselves become primary data objects.

At the same time, unsupervised data-driven techniques, especially principal component analysis (PCA), have emerged as powerful tools to identify dominant fluctuation modes in correlated systems without imposing predefined order parameters \cite{Leiwang2016, Hu2017, Costa2017, Khatami2019, Kiwata2019}. In lattice fermion systems, PCA has been shown to recover magnetic and charge instabilities and to mirror the singular-value structure of generalized susceptibilities \cite{Fahadpca2025, Otsuki2019}. Nevertheless, PCA has rarely been applied systematically to exact real-space correlation matrices of Hubbard clusters across interaction-driven crossovers, nor has its relationship to conventional structure factors been explicitly demonstrated in small geometries. Clarifying this correspondence is essential to assess whether PCA yields genuinely new physical insight or instead provides a systematic and automated reformulation of traditional correlation analysis.

A complementary viewpoint is provided by quantum entanglement measures, which have emerged as sensitive probes of correlation-driven reorganization in many-body systems. Bipartite entanglement entropy has been widely used to characterize interaction-induced localization, quantum criticality, and the buildup of short-range magnetic correlations in Hubbard-like models \cite{Zanardi2002, Amico2008, Zhu2008, Coe2010, Oskar2017,Peled2021, Pontus2022}. In finite clusters, entanglement entropy provides direct access to how quantum information is redistributed between subsystems as charge fluctuations are suppressed and local moments form. Yet, systematic comparisons between entanglement diagnostics, conventional correlation functions, and data-driven PCA in the context of the Mott crossover remain scarce, especially in geometries where spin-sector constraints play a nontrivial role.

A further emerging diagnostic is the quantum-geometry-based distance matrix, which quantifies how rapidly many-body wavefunctions reorganize under parameter changes \cite{Hassan2018, Angelo2020, Yu2025}. This wavefunction-based metric has proven effective in detecting correlation-driven reorganizations in interacting systems \cite{Pbs2014, Hassan2019, Jiang2025}, yet its physical content in the context of the Mott crossover and its relation to conventional correlation measures remains largely unexplored.

Motivated by these developments, we present a multi-diagnostic study of correlation-driven crossovers in finite-size Hubbard clusters, combining conventional observables, real-space correlation functions, entanglement measures, PCA of correlation matrices and quantum-geometry-based distance matrices. Our goals are threefold: (i) to assess the robustness of correlation-driven crossovers across different cluster geometries and boundary conditions, directly addressing finite-size and geometry dependence; (ii) to clarify the role of spin-sector selection in finite systems, where ground-state may carry nonzero total-spin, and to compare half-filled and doped manifolds; (iii) to examine how modern data-driven and wavefunction-based diagnostics complement conventional correlation-function analysis and what additional physical insight they provide.

In this work, we employ ED to investigate a $3\times 3$ square cluster with open boundary conditions (OBC) at half-filling in $S=1/2$ and $S=3/2$ sectors and with one-hole doping in $S=0$ and $S=1$ sectors. We further study a $2\times 4$ cluster with OBC and with mixed i.e., periodic boundary conditions (PBC) along x-direction and OBC along y-direction (cylindrical geometry) at half-filling in $S=0$ and $S=1$ sectors. By considering clusters with both singlet and finite-spin ground-state, we investigate how spin-sector constraints shape correlation diagnostics and test the generality of results obtained from a single finite-geometry \cite{Katnelson2000, Maier2005}.

This paper is organized as follows: In Section \ref{model}, we describe the Hubbard model and computational framework. Section \ref{resdis} presents our results, beginning with order parameter analysis, followed by PCA of charge and spin correlations, and concluding with quantum-geometry-based distance matrices. Finally, Section \ref{summary} summarizes our key findings and their implications for understanding correlation-driven crossovers in finite-size quantum systems.

\section{Model and Method}
\label{model}
The Hamiltonian for one-band Hubbard model \cite{Hubbard1963, Gutzwiller1963, Kanamori1963} in real space is written as 
\begin{equation}
H=-t\sum_{\langle ij\rangle\sigma}(c_{i\sigma}^\dagger c_{j\sigma}+c_{j\sigma}^\dagger c_{i\sigma})+U\sum_i n_{i\uparrow}n_{i\downarrow},
\label{hamil}
\end{equation}
\noindent
where $c_{i\sigma}^\dagger (c_{i\sigma})$ is the fermionic operator that creates (annihilates) an electron with spin $\sigma \in \{\uparrow, \downarrow\}$ at lattice site $i$, and $\langle ij\rangle$ denotes nearest neighbors (NN) sites on the lattice. The parameters $t$ and $U$ represent the NN hopping matrix amplitude and the on-site Coulomb interaction respectively. The number operator $n_{i\sigma} = c_{i\sigma}^\dagger c_{i\sigma}$ counts particles at site $i$ with spin $\sigma$. The tight-binding dispersion relation corresponding to the non-interacting $(U=0)$ case in 2D is given by $\epsilon_k = -2t(cosk_x + cosk_y)$, with a bandwidth $W=8t$. Throughout this work we set $t=1$ as the unit of energy.     

We employ exact diagonalization (ED) to compute ground and low-lying excited states on finite-size clusters. ED provides numerically exact access to eigenvalues, eigenvectors, and arbitrary correlation functions, making it an ideal platform for exploring both conventional observables and emerging data-driven or wavefunction-based diagnostics. To extend the accessible system size and resolve different total-spin manifolds explicitly, we exploit full spin-rotational symmetry and construct spin-adapted basis states \cite{Sarmahsan1996}. This symmetry adaptation block-diagonalizes the Hamiltonian into fixed-spin sectors and significantly reduces the effective Hilbert-space dimensionality, allowing us to investigate specific spin manifolds with high numerical precision. The use of spin-adapted ED is essential for systematically examining spin-sector dependence of correlation-driven crossovers in finite systems.

While ED is restricted to small lattices by the exponential growth of the Hilbert space dimensionality, small Hubbard clusters remain valuable theoretical laboratories: they retain the full local interaction structure of the model, capture the short-range spin and charge correlations that dominate the Mott crossover, and permit controlled exploration of geometry-, boundary-, and spin-sector dependence. This is particularly relevant in light of recent discussions on the representability of finite-cluster results for Hubbard model physics and the increasing experimental realization of engineered small correlated systems in quantum simulators and nanostructures.

In this work, we consider multiple cluster geometries and boundary conditions to assess the robustness of correlation-driven Mott crossovers beyond a single finite-size realization. Specifically, we study a $3\times3$ cluster with open boundaries at half-filling in $S=1/2$ and $S=3/2$ sectors and with one-hole doping in $S=0$ and $S=1$ sectors, as well as a $2\times4$ cluster with open and cylindrical boundary conditions at half filling in $S=0$ and $S=1$ sectors. This ensemble includes both singlet and finite-spin ground-state, enabling us to examine how spin-sector constraints and cluster geometry influence correlation functions, entanglement measures, PCA modes and quantum-geometry-based distance metrics. Such a comparative finite-size analysis allows us to distinguish generic correlation-driven trends from geometry-specific artifacts, thereby providing a controlled platform for evaluating the physical content and practical utility of modern data-driven diagnostics in strongly correlated electron systems.

\section{Results and Discussion}
\label{resdis}
In this section, we present an analysis of correlation-driven crossovers in finite-size Hubbard clusters using several complementary diagnostics. We begin with the analysis of conventional order-parameters to establish baseline signatures of the interaction-driven metal-insulator crossover across different geometries and spin sectors. We then turn to PCA of real-space charge and spin correlation matrices to uncover dominant fluctuation modes and assess their correspondence with traditional structure-factor-based diagnostics. Finally, we employ quantum-geometry-based distance matrices to characterize wavefunction reorganization across the Mott crossover. This progression from standard observables to data-driven and wavefunction-based measures allows us to systematically benchmark emerging diagnostics along with established correlation-function analysis.

\subsection{Order parameter investigations}
\subsubsection{Spin-resolved charge gap}
An important signature of the Mott crossover is the opening of a charge (Mott) gap, which measures the energy cost to change the particle number in the system by one. This gap provides a diagnostic measure of interaction-driven incompressibility and suppression of charge fluctuations. However, in small clusters, the lowest-energy states in neighboring particle-number sectors may belong to different total-spin manifolds. To examine the interplay between the spin and the charge degrees of freedom leading to Mott crossover in such finite systems, we compute the spin-resolved charge gap as defined in the following. Denoting the ground-state energy $E_0(N, S_N^L)$ for $N$ electrons with the lowest total-spin $S_N^L$ (where $S_N^L=0$ for even $N$ and $S_N^L=1/2$ for odd $N$), and the next higher-spin channel as $S_N^H=S_N^L +1$, the charge gap is defined as        
\begin{equation}
\Delta_c(S_N^L) = E_0(N+1, S_{N+1}^L) + E_0(N-1, S_{N-1}^L) - 2E_0(N, S_N^L).
\end{equation}
This construction allows us to track how spin correlations reorganize upon particle addition or removal across the interaction-driven crossover.

\begin{figure}[h]
\centering
\includegraphics[scale=0.60]{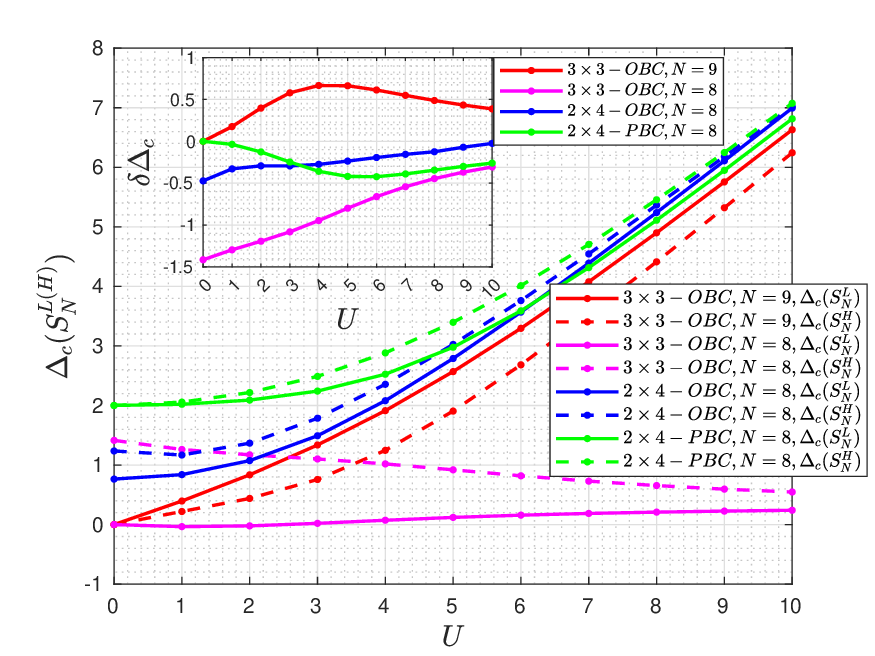}
\caption{Spin-resolved charge-gaps $\Delta_c(S_N^{L(H)})$ as a function of interaction strength $U$ for different cluster geometries, fillings, and boundary conditions. Solid and dashed lines denote the lowest-spin and the next higher-spin excitation channels, respectively. Half-filled clusters ($N=9$ for $3\times3$ and $N=8$ for $2\times4$) exhibit a monotonic opening of the spin-resolved charge-gaps with increasing $U$, signaling the onset of interaction-driven incompressibility. In contrast, the one-hole-doped $3\times3$ cluster ($N=8$) exhibits nearly zero charge-gap corresponding to the lowest-spin channel over the full interaction range, reflecting finite compressibility away from half-filling. Inset: Difference between the charge-gaps corresponding to the lowest-spin and the next higher-spin channels, $\delta\Delta_c=\Delta_c(S_N^L)-\Delta_c(S_N^H)$, which peaks at intermediate coupling and approaches the effective exchange scale $J\sim 4t^2/U$ at large $U$, indicating residual spin-dependent energy splitting of charge excitations.}
\label{fig:chargegap}
\end{figure}

In Fig.~\ref{fig:chargegap} we display the evolution of the spin-resolved charge gaps $\Delta_c(S_N^L)$ and $\Delta_c(S_N^H)$, for the lowest-spin channel and the next higher-spin channel respectively, as function of interaction strength $U$ for various cluster geometries and boundary conditions. For the half-filled clusters, both $\Delta_c(S_N^L)$ and $\Delta_c(S_N^H)$ increase monotonically with $U$, reflecting the progressive suppression of charge fluctuations and emergence of interaction-driven incompressibility, characteristic of the Mott crossover. The gap increases rapidly beyond weak coupling regime, demonstrating that even small clusters capture the essential short-range physics of Mott localization.
A key observation is that the magnitude of the spin-resolved charge gaps $\Delta_c(S_N^L)$ and $\Delta_c(S_N^H)$ depends sensitively on the total-spin $S_N^L$ and $S_N^H$ of the underlying ground-state. This can be understood physically as follows. Adding or removing an electron drives the system into a different total-spin sector, and the associated energy cost is governed by how strongly the resulting spin configuration changes the spin correlations of the original $N$-particle ground state.
In the $3\times3$ half-filled cluster, the charge gap in the lowest-spin channel remains more strongly gapped than the next higher-spin channel across the entire interaction range, i.e., $\Delta_c(S_N^L) > \Delta_c(S_N^H)$. 
The charge excitation in $N=9$ electrons on $3\times 3$ lattice connects the lowest-spin $S_N^L=1/2$ ground-state to the lowest-spin states in the $N\pm1=10,8$ particle-sectors.
The next higher-spin state with $S_N^H=3/2$ having three unpaired electrons undergoes reorganization in its spin correlation to accommodate the added or removed elecron resulting in charge-excited next higher-spin state with $S_{N\pm1}^H=1$ having two unpaired electrons thus incurring a lower energy cost for doublon-holon creation \cite{Okamoto2019} compared to lowest-spin channel.
In contrast, for the $2\times4$ half-filled cluster with singlet ground-state ($S_N^L=0$), the hierarchy of spin-resolved charge gap is reversed: the lowest-spin charge gap is lower than the next higher-spin charge gap i.e., $\Delta_c(S_N^L) < \Delta_c(S_N^H)$. 
In the antiferromagnetically ordered half-filled $2\times 4$ cluster with the lowest-spin $S_N^L=0$, the charge-excitation involves going into $N\pm1$ sector in the lowest-spin $S_{N\pm1}^L=1/2$ state having one unpaired electron results in small charge gap for the lowest-spin channel. 
In the next higher-spin channel $S_N^H=1$ with two unpaired electrons, the charge-excitation takes the system into $N\pm1$ particle-sector in the next higher-spin $S_{N\pm1}^H=3/2$ state with three unpaired electrons thereby moving the system further away from antiferromagnetic spin ordering.  
Thus, the spin $S_N^{L(H)}$ of the reference ground-state provides a sensitive measure of the corresponding spin-resolved charge gaps $\Delta_c(S_N^L)$ and $\Delta_c(S_N^H)$.  
The preceeding discussion on the two clusters $3\times 3$ and $2\times 4$  considered in the present study leads us to conclude that at half-filling the absolute magnitude of the spin-resolved charge-gaps increase with $U$, confirming that the Mott gap itself is the generic feature, while its spin-resolved counterpart reveals finite-size and spin-sector constraints.

In one-hole-doped $3\times3$ cluster with $N=8$ electrons, the spin-resolved charge-gaps $\Delta_c(S_N^L)$ and $\Delta_c(S_N^H)$ in the lowest-spin channel ($S_N^L, S_{N\pm1}^L$)=(0,1/2) and the next higher-spin channel ($S_N^H, S_{N\pm1}^H$)=(1,3/2) remain small but reveal distinct behaviors about spin-charge coupling. 
In the lowest-spin channel, the charge excitation involves system going from $N=8$ to $N\pm1=9,7$ particle-sectors with charge gap $\Delta_c(S_N^L)$ becoming zero for weak to intermediate coupling and developing a minimal value at strong $U$, reflecting the fact that the one-hole doped in half-filled Hubbard cluster restores compressibility even at strong coupling. 
This can be understood as, once the system is doped away from half-filling, charge excitations no longer require creating or removing a doublon-holon pair \cite{Okamoto2019} across the Mott gap, instead low-energy particle-number fluctuations correspond to redistribution of the mobile hole, which costs only kinetic energy and remains finite even for large $U$. Consequently, the doped cluster behaves as a correlated metal, exhibiting finite compressibility and the absence of a true Mott gap.
In contrast, the charge excitation in the next higher-spin channel takes the system from the spin state $S_N^H=1$ with two unpaired electrons to $S_{N\pm1}^H=3/2$ with three unpaired electrons in the neighboring particle-sectors yielding a finite charge gap $\Delta_c(S_N^H)$ even at $U=0$ which arises from the kinetic and exchange energies required to reconfigure electrons into the next higher-spin multiplet. As $U$ increases, $\Delta_c(S_N^H)$ decreases, signaling that strong correlation suppressess double occupancy and enhances coherent hole motion within the background spin configuration, thereby reducing the kinetic energy cost of accommodating the additional unpaired spin. 

The inset of Fig. \ref{fig:chargegap} shows the difference between the charge gaps corresponding to the lowest-spin and the next higher-spin channels, $\delta\Delta_c=\Delta_c(S_N^L)-\Delta_c(S_N^H)$, which provides a measure of how addition or removal of charge reorganizes the spin correlations within the system.  
We observe that, both the sign and the magnitude of $\delta\Delta_c$ depend on the spin of the reference ground state and whether the system is half-filled or hole-doped. 
For the half-filled $3\times 3$ cluster, $\delta\Delta_c$ starts from zero, rises to a pronounced positive maximum at intermediate $U$, and then decreases at large $U$. The initial growth in $\delta\Delta_c$ at intermediate $U$ values reflects rapid development of short-range singlet correlations. 
At large $U$, the decrease of $\delta\Delta_c$ implies that the spin-dependent part of the charge gap is controlled by the effective exchange energy scale $J\sim 4t^2/U$, while the dominant charge excitation energy is determined by $U$. This confirms that in the strong-coupling limit the splitting between spin-resolved charge gaps is governed by $J$. 
In the one-hole-doped $3\times 3$ cluster, $\delta\Delta_c$ is negative maximum at $U=0$ and increases gradually toward zero with increasing $U$. The negative sign indicates that in comparison to the lowest-spin charge excitations the next higher-spin charge excitations costs more energy when a mobile hole is present. As interactions grow, the motion of the hole becomes increasingly constrained by short-range antiferromagnetic correlations, but the effective spin exchange scale $J$ simultaneously decreases as $1/U$. The gradual reduction in the magnitude of $\delta\Delta_c$ therefore reflects the diminishing influence of spin-dependent exchange processes on predominantly kinetic, compressible charge excitations in the one-hole-doped system.
For the half-filled $2\times 4$ cluster with open boundaries, $\delta\Delta_c$ starts negative, contrasting with the $3\times 3$ half-filled cluster. This reflects the singlet ground-state preference for the lowest-spin charge excitations. With increasing $U$ local moment formation suppresses this distinction, and $\delta\Delta_c$ move towards zero, controlled by the decaying $J$ scale.
Finally, for the half-filled $2\times 4$ cluster with cylindrical boundary conditions, $\delta\Delta_c$ shows a non-monotonic evolution: starting near zero, developing a negative minimum at intermediate $U$ and increasing at large $U$. This behavior signals a competition between geometric frustration induced by the boundary conditions at intermediate coupling and the eventual dominance of local short-range correlations at strong coupling.

Taken together, these trends demonstrate that while the opening of the Mott gap itself is robust across geometries, the sign and the magnitude of the spin splitting of the charge gap reflects the spin structure of the underlying ground state and finite-size geometry. In half-filled clusters, the decay of $\mid\delta\Delta_c\mid$ at large $U$ provides direct numerical evidence that the spin-dependent part of the charge gap is governed by the effective exchange energy $J\sim 4t^2/U$.

\subsubsection{Local observables: local moment and double occupancy}
After establishing the opening of spin-dependent charge-gap across the interaction-driven Mott crossover, we now turn to local real-space observables that directly quantify the microscopic mechanisms underlying this crossover. In particular, the suppression of double occupancy reflects the freezing of charge fluctuations, while the growth of local magnetic moments signals the emergence of well-formed spin degrees of freedom. These on-site quantities therefore provide a minimal and physically transparent diagnostic of Mott localization, complementary to the nonlocal charge-gap analysis.

We define the average local moment and double occupancy as
\begin{eqnarray}
\bar{m}=\frac{4}{N}\sum_i\langle (n_{i\uparrow}-n_{i\downarrow})^2\rangle,
\\
\bar{d}=\frac{M}{N^2}\sum_i\langle n_{i\uparrow}n_{i\downarrow}\rangle,
\end{eqnarray}
where $M$ is the number of lattice sites. At half-filling $N=M$. These observables respectively measure the degree of spin polarization per site and the probability of finding doubly occupied sites. Their evolution with interaction strength provides a direct measure of how local degrees of freedom reorganize across the correlation-driven crossover.
\begin{figure}[h]
\centering
\includegraphics[scale=0.60]{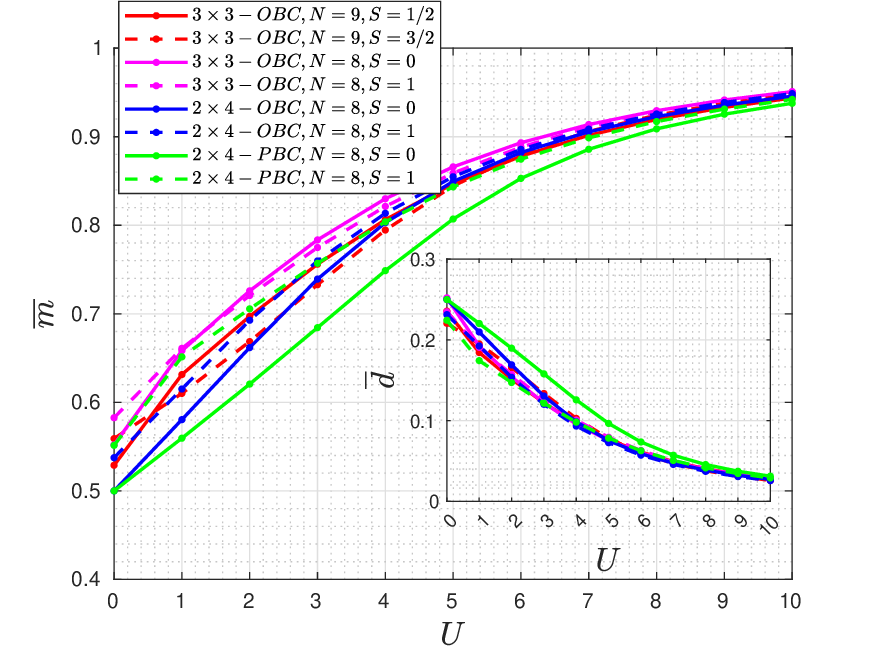}
\caption{Average local moment $\bar{m}$ (main panel) and average double occupancy $\bar{d}$ (inset) as functions of interaction strength $U$ for all cluster geometries, fillings, and spin sectors studied. The monotonic growth of $\bar{m}$ and suppression of $\bar{d}$ reflect the progressive formation of local moments and freezing of charge fluctuations across the Mott crossover.}
\label{fig:lclmdbc}
\end{figure}

Figure \ref{fig:lclmdbc} displays the evolution of the average local moment $\bar{m}$ (main panel) and average double occupancy $\bar{d}$ (inset) for all cluster geometries, fillings, and spin sectors considered. In half-filled clusters, $\bar{d}$ decreases rapidly with increasing $U$, while $\bar{m}$ grows toward saturation. This complementary behavior directly reflects the suppression of on-site charge fluctuations and the concomitant formation of well-defined local moments --- the defining microscopic mechanism of Mott localization. The most rapid variation occurs in the weak-to-intermediate coupling regime, coinciding with the region where the charge-gap opens most strongly, confirming that the crossover identified in the charge-sector analysis is driven by the freezing of local charge dynamics. 
The one-hole-doped $3\times 3$ cluster displays a distinct trend. Although $\bar{d}$ is still suppressed with increasing $U$, its magnitude remains systematically larger than in the half-filled clusters, and the local moment grows more slowly. This reflects the persistence of mobile charge carriers that restore finite compressibility and weaken local moment formation, consistent with the small charge-gap found in the one-hole-doped case. The residual growth of $\bar{m}$ at large $U$ indicates that strong local moments still form on most sites, but their coupling to the itinerant hole prevents full saturation.

Taken together, these results confirm that the local observables provide a consistent microscopic interpretation of the charge-gap analysis: half-filled clusters undergo a correlation-driven crossover into a local-moment regime, while doped cluster remain compressible with partially suppressed moment formation. Moreover, the comparison across geometries and spin sectors demonstrates that while finite-size effects influence weak-coupling behavior, the strong-coupling local physics is universal.

\subsubsection{Charge and spin correlations}
To directly probe how electronic correlations reorganize in real space across the interaction-driven Mott crossover, we next examine two-point charge-charge and spin-spin correlation functions. While local observables such as double occupancy and local moments capture on-site physics, two-point correlators reveal how fluctuations propagate between different lattice sites and therefore provide a microscopic window into the development of nonlocal charge coherence and magnetic ordering tendencies. We define the charge-charge and spin-spin correlation functions as 
\begin{eqnarray}
D_{ij}=\langle G\mid n_i n_j\mid G\rangle, 
\\
L_{ij}=\frac{1}{4}\langle G\mid (n_{i\uparrow}-n_{i\downarrow})(n_{j\uparrow}-n_{j\downarrow}) \mid G\rangle,
\end{eqnarray}
where $\mid G\rangle$ is the ground state and $n_i=n_{i\uparrow}+n_{i\downarrow}$. As the number of lattice sites increases, so does the complexity of the correlation functions between different sites. Therefore, it is useful to define structure factors that translate these correlations into the momentum space. The charge and spin structure factors are defined as
\begin{eqnarray}
S_c(\mathbf{q})=\frac{1}{M}\sum_{ij} e^{i\mathbf{q}\cdot(\mathbf{R_i}-\mathbf{R_j})}D_{ij},
\\
S_s(\mathbf{q})=\frac{1}{M}\sum_{ij} e^{i\mathbf{q}\cdot(\mathbf{R_i}-\mathbf{R_j})}L_{ij},
\end{eqnarray}
where $\mathbf{q}$ is the wave vector and $\mathbf{R_i(R_j)}$ denote the positions of lattice site $i(j)$. We focus on $\mathbf{q}=(\pi,\pi)$, which is relevant for probing correlations at the antiferromagnetic wave vector. 
\begin{figure}[h]
\centering
\includegraphics[scale=0.60]{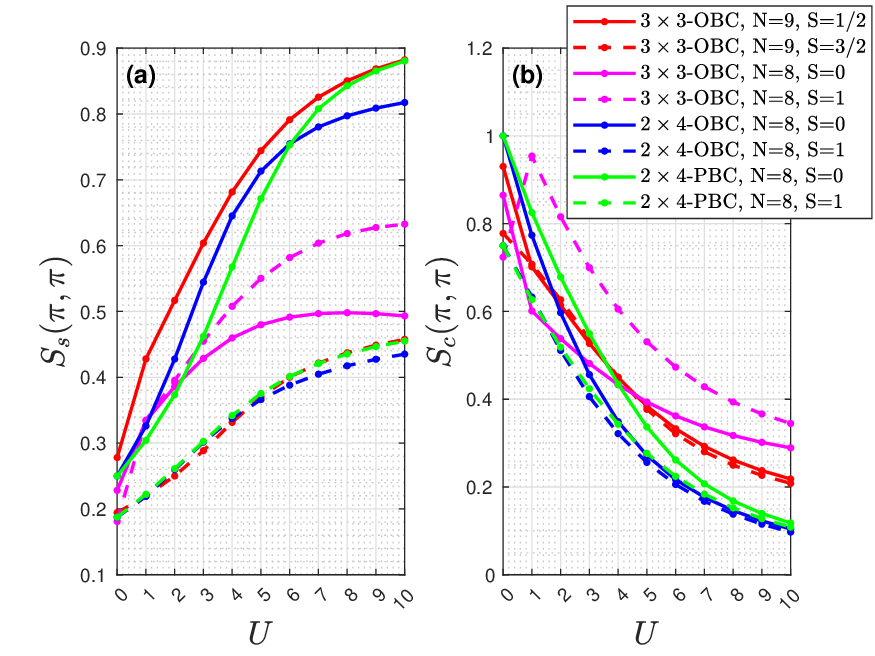}
\caption{Spin and charge structure factors at the antiferromagnetic wave vector $\mathbf{q}=(\pi,\pi)$ as functions of interaction strength $U$ for all cluster geometries, fillings, and spin sectors studied. (a) Spin structure factor $S_s(\pi,\pi)$. (b) Charge structure factor $S_c(\pi,\pi)$. The growth of $S_s(\pi,\pi)$ and suppression of $S_c(\pi,\pi)$ with increasing $U$ reflect the development of short-range antiferromagnetic correlations and the freezing of charge fluctuations across the Mott crossover.}
\label{fig:sfs}
\end{figure}

In Fig. \ref{fig:sfs} we present the evolution of the spin and charge structure factors at the antiferromagnetic wave vector $\mathbf{q}=(\pi,\pi)$, providing a direct measure of nonlocal correlation buildup across the interaction-driven crossover. In half-filled clusters, the spin structure factor $S_s(\pi,\pi)$ grows monotonically with increasing $U$, signaling the progressive formation of short-range antiferromagnetic correlations as local moments develop. Conversely, the charge structure factor $S_c(\pi,\pi)$ is strongly suppressed with increasing $U$, reflecting the freezing of intersite charge fluctuations accompanying Mott localization. The simultaneous enhancement of spin correlations and depletion of charge coherence thus provides a conventional real-space confirmation of the crossover already identified by charge gaps and local observables.
Quantitative differences across geometries and spin sectors reveal finite-size effects and ground-state spin constraints. Nevertheless, the half-filled geometries exhibit the same qualitative trend, confirming that the buildup of short-range antiferromagnetism is robust and not an artifact of a single cluster.
The one-hole-doped $3\times 3$ cluster behaves distinctly. Here, $S_s(\pi,\pi)$ still increases with $U$, indicating that strong local moments form even in the presence of a mobile hole. However, its saturation value remains substantially lower than in half-filled clusters, reflecting the disruption of antiferromagnetic order by hole motion. Simultaneously, $S_c(\pi,\pi)$ remains significantly larger than in half-filled cases and decreases only slowly with $U$, consistent with the persistence of low-energy charge fluctuations and the near-vanishing charge gap observed in the doped system. This directly links the absence of a Mott gap in the doped cluster to the survival of finite intersite charge coherence.

We observe that, these structure-factor results establish a direct correspondence between conventional correlation diagnostics and the PCA analysis presented later: the dominant PCA modes extracted from correlation matrices correspond precisely to the real-space patterns underlying $S_s(\pi,\pi)$ and $S_c(\pi,\pi)$. Thus, PCA does not merely reproduce known information but provides an automated and compact representation of these correlation channels.

\subsubsection{von-Neumann entanglement entropy}
To characterize how interaction-driven localization reorganizes global many-body quantum correlations, we next examine the bipartite von-Neumann entanglement entropy of the exact ground state. Unlike local observables or two-point correlators, entanglement entropy provides a basis-independent measure of how quantum information is shared across the entire system, and is therefore particularly sensitive to interaction-driven changes in the structure of the many-body wave function.

We obtain the ground state $\mid \psi_0\rangle$ in each relevant total-spin sector and partition the lattice into two complementary sublattices A and B. The reduced density matrix of subsystem A is computed by tracing over the degrees of freedom of B
\begin{equation}
\rho_A=Tr_B(|\psi_0\rangle\langle\psi_0|),
\end{equation}
and the entanglement entropy is then defined as
\begin{equation}
S_A=-Tr(\rho_A log_2 \rho_A).
\end{equation}
This quantity directly probes how the effective Hilbert-space connectivity between spatial regions evolves as charge fluctuations are suppressed and spin correlations become dominant across the Mott crossover.

\begin{figure}[h]
\centering
\includegraphics[scale=0.60]{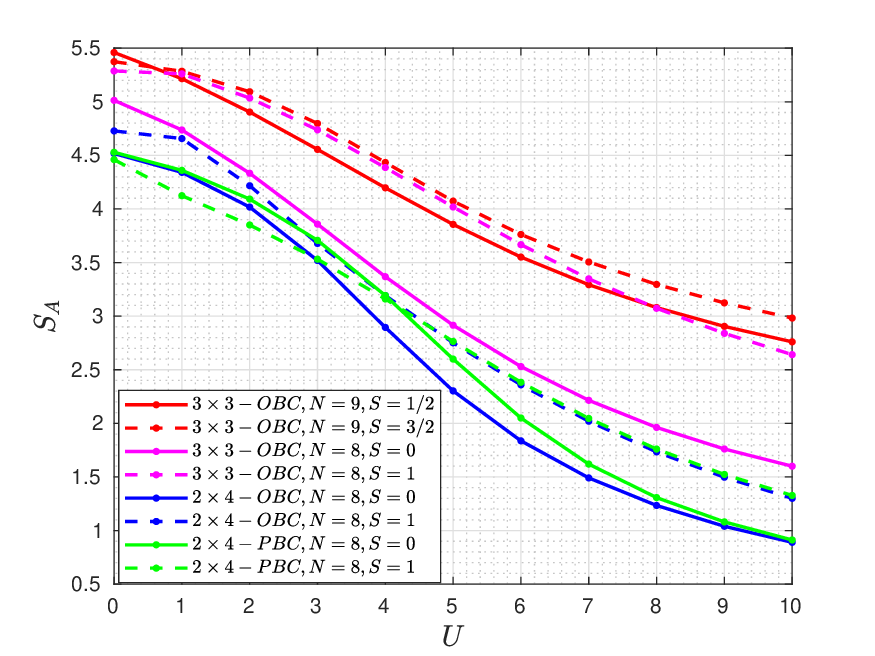}
\caption{Bipartite von-Neumann entanglement entropy $S_A$ versus interaction strength $U$ for all cluster geometries, fillings, and spin sectors studied. The systematic reduction of entanglement with increasing $U$ reflects the progressive localization of electrons and the simplification of the many-body wave function across the Mott crossover.}
\label{fig:vne}
\end{figure}

\begin{figure*}[!htbp]
\centering
\includegraphics[scale=0.36]{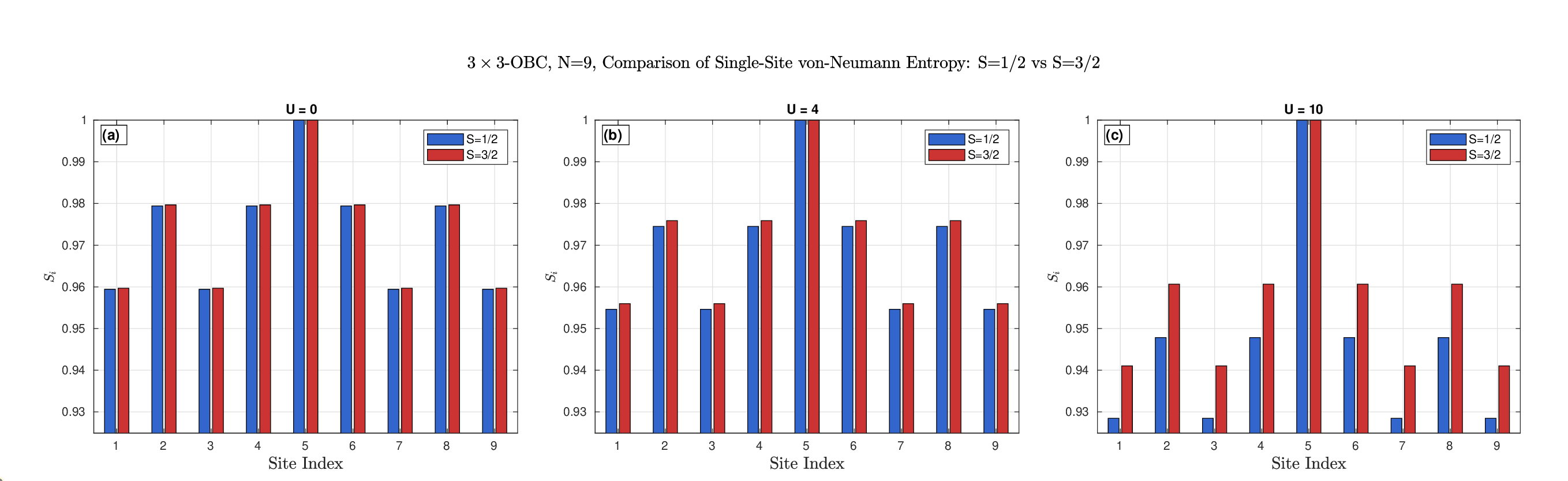}
\quad
\includegraphics[scale=0.36]{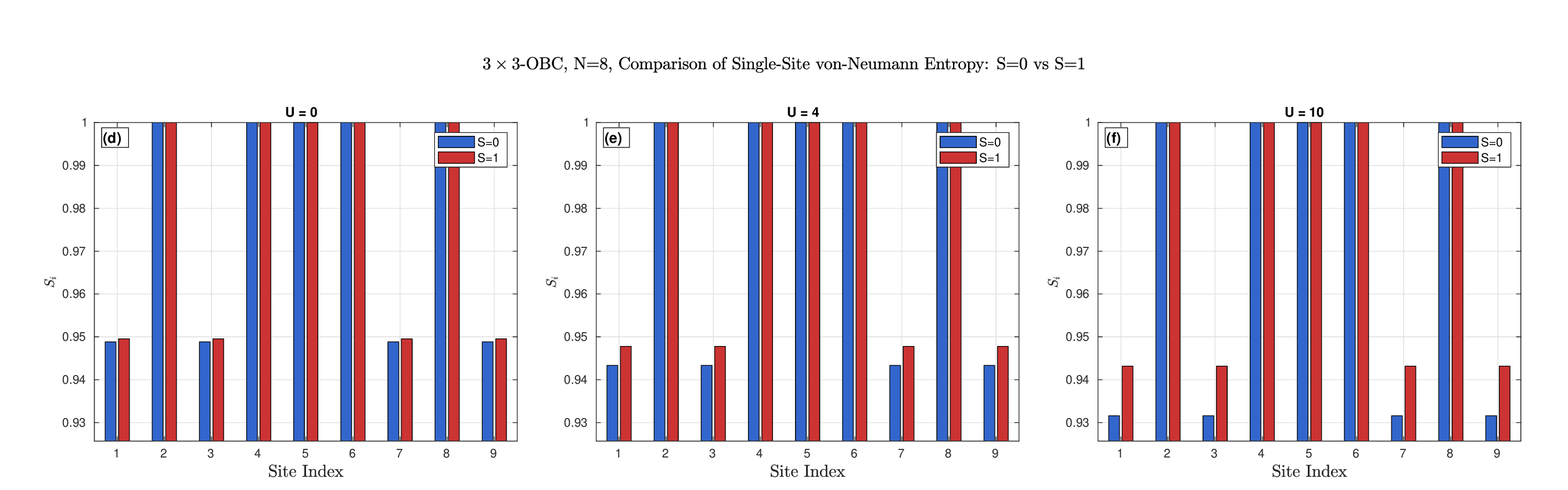}
\quad
\includegraphics[scale=0.36]{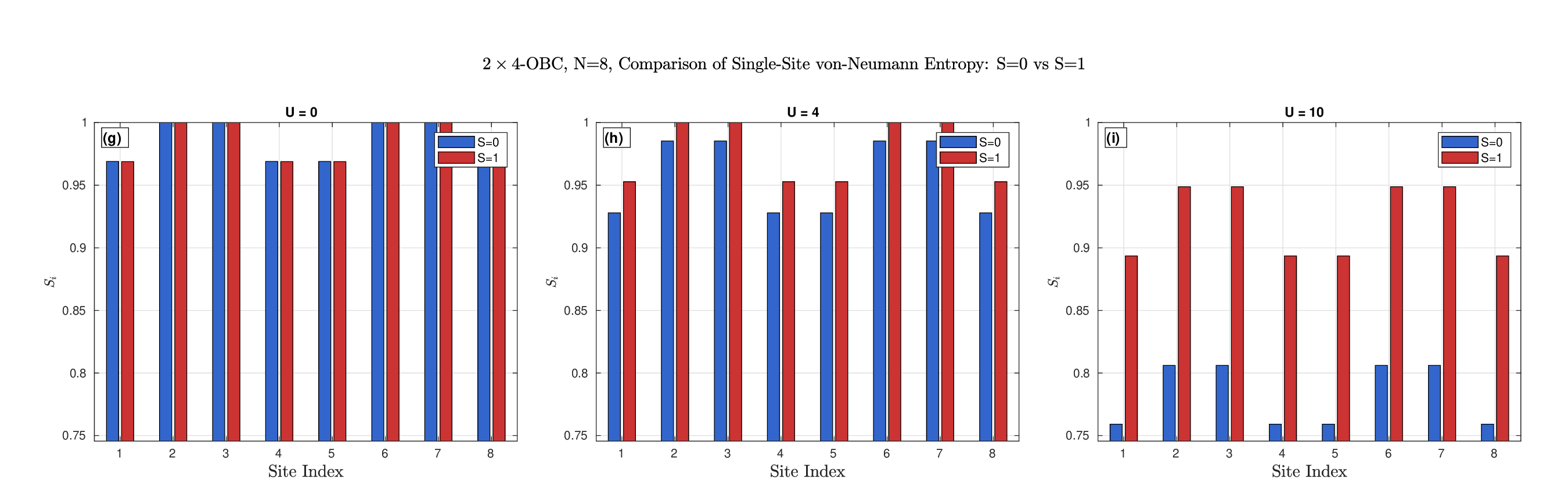}
\quad
\includegraphics[scale=0.36]{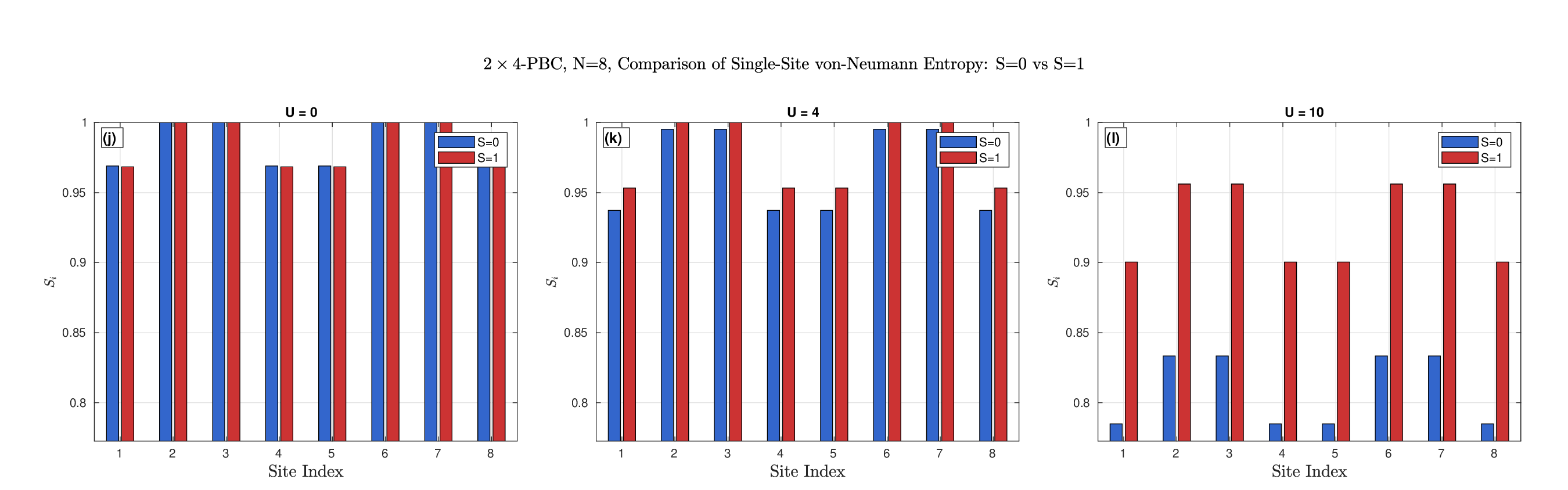}
\caption{Single-site von-Neumann entropy $S_i$ for different cluster geometries and spin sectors at representative interaction strengths $U=0,4,10$. (a-c) $3\times 3$ cluster at half-filling ($N=9$) comparing $S=1/2$ and $S=3/2$ sectors. (d-f) $3\times 3$ cluster with one hole ($N=8$) comparing $S=0$ and $S=1$ sectors. (g-i) $2\times 4$ cluster with open boundaries at half-filling. (j-l) $2\times 4$ cluster with cylindrical boundaries at half-filling. Bars show site-resolved entropy; edge and corner sites exhibit reduced entanglement compared to bulk sites, and higher-spin sectors systematically display enhanced local entanglement at strong coupling.}
\label{fig:ssvne}
\end{figure*}
Figure \ref{fig:vne} shows the evolution of the bipartite entanglement entropy as a function of interaction strength. In the half-filled clusters, $S_A$ decreases monotonically with increasing $U$, demonstrating that the many-body ground state becomes progressively less entangled as charge fluctuations are frozen and electrons localize. In the non-interacting limit, the wave function is highly delocalized and entanglement is maximal. As $U$ grows, the suppression of double occupancy truncates the accessible Hilbert space, leading to a systematic reduction of spatial entanglement.
Notably, half-filled clusters with singlet ground states (the $2\times 4$ geometries) exhibit a faster decay and lower saturation value of $S_A$ than the $3\times 3$ cluster, whose ground state carries a finite total spin. This reflects the fact that an uncompensated spin necessarily maintains residual long-range quantum correlations between subsystems, preventing full disentanglement even deep in the strong-coupling regime. This provides an entanglement-based explanation of the reduced spin structure factor saturation and smaller low-spin charge gap observed earlier in the $3\times 3$ cluster.
The one-hole-doped $3\times 3$ cluster displays qualitatively different behavior. Although $S_A$ still decreases with $U$, it remains substantially larger than in the $2\times 4$ half-filled cases even at strong coupling. This persistence of finite entanglement originates from the mobile doped hole, which continues to delocalize across the system and entangles distant regions despite strong on-site repulsion. This entanglement signature therefore provides a direct, basis-independent confirmation that the doped cluster remains compressible and metallic, in agreement with the near-vanishing charge gap and finite charge structure factor reported earlier.
A further important observation emerges at intermediate coupling: the rate of decrease of $S_A$ changes around $U\sim 4-5$, coinciding with the maximum in the spin-dependent charge-gap separation and the rapid growth of short-range antiferromagnetic correlations. This identifies the crossover scale directly in the entanglement structure of the wave function, demonstrating that entanglement entropy provides an independent and highly sensitive diagnostic of the interaction-driven reorganization of spin and charge degrees of freedom.

Finally, in the large $U$ limit, the residual entanglement saturates at a value governed by spin exchange processes alone. This is consistent with the effective Heisenberg description, where the only remaining quantum correlations arise from effective exchange interactions of scale $J\sim 4t^2/U$.

The single-site von-Neumann entropy, $S_i=-Tr(\rho_i log_2 \rho_i)$, where $\rho_i$ is the reduced density matrix of site $i$, provides a strictly local measure of quantum entanglement between an individual lattice site and the rest of the system. Unlike the bipartite entropy, which captures global entanglement across an extended spatial cut, $S_i$ directly quantifies how local charge and spin degrees of freedom become entangled with surrounding sites. It therefore offers an intuitive real-space visualization of how interaction-driven suppression of charge fluctuations and the buildup of local moments reorganize quantum correlations at the single-site level. In finite clusters, where translational invariance is broken by geometry and boundaries, the site-resolved entropy further reveals how edge, corner, and bulk sites participate differently in the correlation-driven crossover.

Figure \ref{fig:ssvne} displays the single-site von-Neumann entropy across cluster geometries, boundary conditions, and spin sectors. In the noninteracting limit $U=0$, every site has a large single-site entropy $S_i$, demonstrating that the ground state is highly mixed on each site and that charge and spin degrees of freedom are delocalized across the cluster. This is the expected metallic-like behaviour. As $U$ increases, $S_i$ decreases at all sites, signaling the progressive suppression of local charge fluctuations and the emergence of well-formed local moments. This reduction is spatially nonuniform: corner and edge sites exhibit systematically lower entropy than interior sites, indicating that reduced coordination at boundaries enhances localization and weakens entanglement with the rest of the cluster.
A clear spin-sector dependence is visible at strong coupling. For half-filled clusters, higher-spin sectors display systematically larger single-site entropies than lower-spin sectors. This reflects that polarized or partially polarized backgrounds allow greater local spin uncertainty once charge degrees of freedom are frozen, whereas singlet-dominated backgrounds lock spins into more rigid short-range correlations, reducing local entanglement. This trend is consistent with the spin-resolved charge-gaps analysis, where the lowest-spin channel exhibited large charge-gap and therefore more strongly suppressed local fluctuations. 
In doped clusters, the site entropy remains higher and shows weaker suppression with increasing $U$, confirming that mobile holes restore local charge fluctuations and maintain finite compressibility even at strong coupling. This local entanglement perspective thus complements the charge-gap results, providing a real-space visualization of how doping melts the Mott localization tendency in finite systems.
Finally, comparing open and cylindrical boundaries in the $2\times 4$ clusters shows that boundary conditions primarily affect the spatial distribution of $S_i$ but not the overall interaction-driven trend, confirming that the observed entanglement restructuring is a robust correlation effect rather than the artifact of boundary conditions. 

\subsection{PCA of charge and spin correlations}
To obtain an unbiased, data-driven characterization of how correlation patterns reorganize across the interaction-driven crossover, we apply principal component analysis (PCA) directly to the real-space charge-charge correlation ($D_{ij}$) and spin-spin correlation ($L_{ij}$) matrices introduced in the previous section. Unlike conventional structure factors, which project correlations onto predetermined momentum channels, PCA identifies the dominant collective fluctuation modes solely from the statistical structure of the correlation data, without imposing any predefined order parameter or wave vector.

The site-resolved correlation matrices $D_{ij}$ and $L_{ij}$ computed at various $U$ values are reshaped into data vectors and centered by subtracting the mean of each component. From the centered data matrix $X$, the covariance matrix is defined as
\begin{equation}
C=\frac{1}{M}X^{T}X,
\end{equation}
which is an $M\times M$ real symmetric matrix. Diagonalization of $C$ yields eigenvalues $\lambda_k$ and orthonormal eigenvectors $w_k$,
\begin{equation}
Cw_k=\lambda_k w_k, \hskip 1cm k=1,...,M.
\end{equation}
Each eigenvector $w_k$ defines a principal component, i.e., a collective fluctuation pattern in correlation space. The associated eigenvalue $\lambda_k$ measures how much of the total variance of the dataset is carried by that fluctuation mode.
The projection of a given correlation dataset $X$ onto the $k$-th principal component is
\begin{equation}
p_k=X\cdot w_k.
\end{equation}
Tracking $p_k$ as a function of $U$ therefore reveals how strongly each collective mode contributes to the correlations at that interaction strength.
To quantify the statistical importance of each principal component, we define the normalized explained variance-ratio
\begin{equation}
\tilde{\lambda}_k = \frac{\lambda_k}{\sum_i \lambda_i}.
\end{equation}
The leading variance-ratios $\tilde{\lambda}_k$ reveal which fluctuation channels dominate the correlation data. By tracking their evolution with the interaction strength $U$, we can identify charge-dominated or spin-dominated regimes in a fully data-driven manner. This unsupervised approach thus complements traditional observables and provides a transparent framework for visualizing correlation reorganization and crossovers in finite-size Hubbard clusters. 

\begin{figure}[h]
\centering
\includegraphics[scale=0.42]{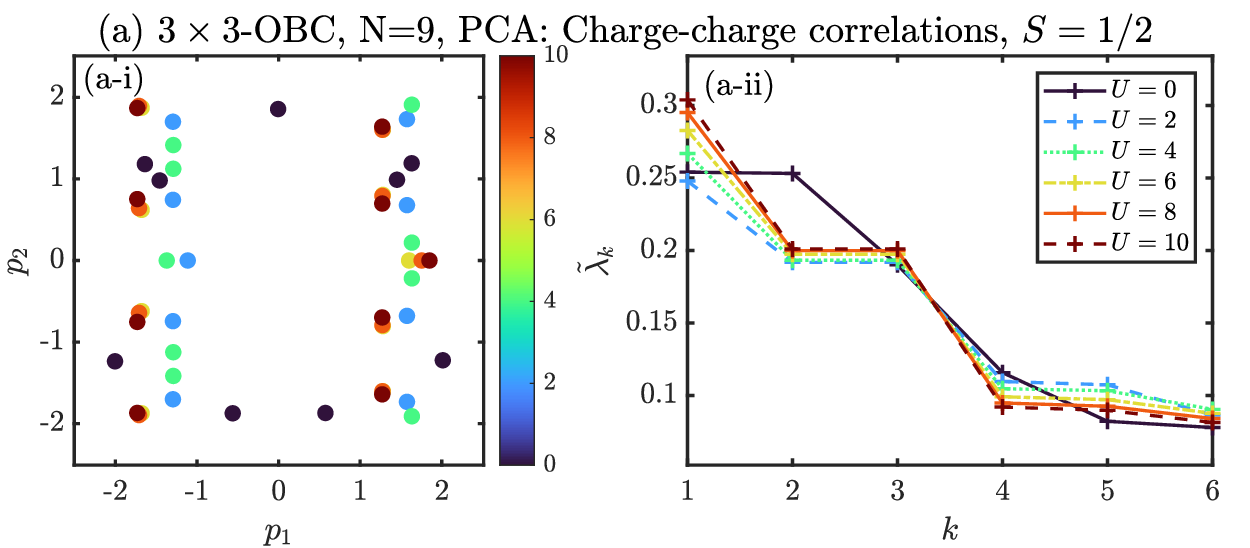}
\quad
\includegraphics[scale=0.42]{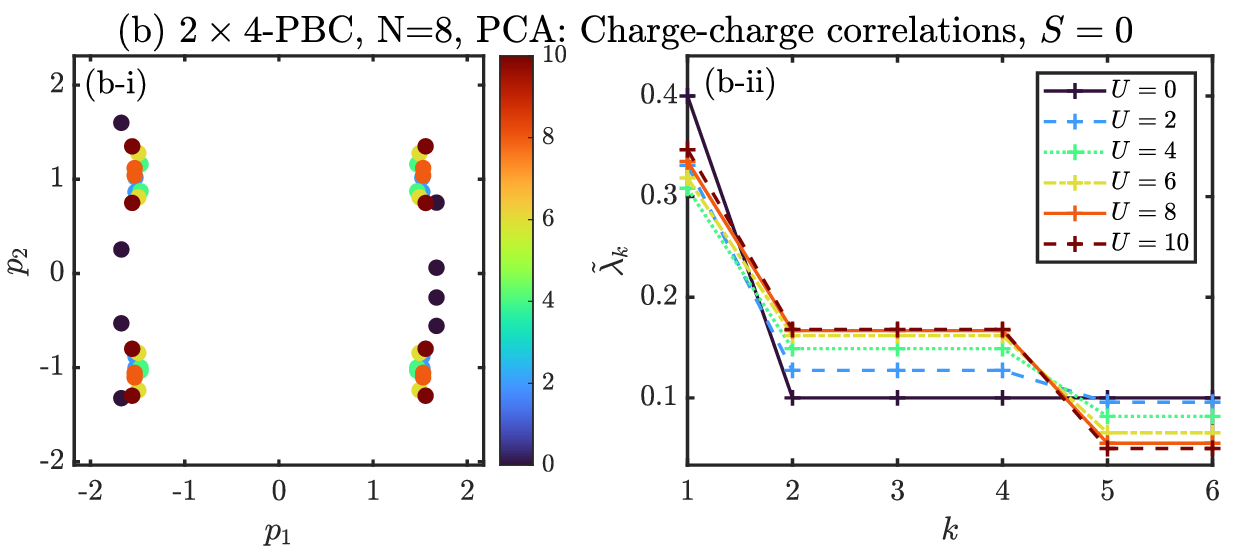}
\caption{Principal component analysis of charge-charge correlation matrices. (a) $3\times 3$ cluster with OBC at half-filling in the $S=1/2$ sector. Panel (a-i) shows the projections of correlation data onto the first two principal components $p_1$ and $p_2$, with color indicating interaction strength $U$. Panel (a-ii) displays the explained variance ratios $\tilde{\lambda}_k$ of the leading six components for several values of $U$. (b) $2\times 4$ cluster with cylindrical boundary conditions at half-filling in the $S=0$ sector. Panels (b-i) and (b-ii) are defined analogously.}
\label{fig:chrgpca}
\end{figure}

In Fig. \ref{fig:chrgpca} we present the PCA of charge-charge correlation matrices for two representative half-filled clusters: the $3\times 3$ OBC system in the $S=1/2$ sector and the $2\times 4$ cylindrical system in the $S=0$ sector. In both cases, the projections onto the first two components $p_1$-$p_2$ form well separated trajectories as $U$ increases, indicating a continuous reorganization of real-space charge correlations across the interaction-driven crossover. The absence of abrupt clustering confirms that the finite clusters undergo a smooth crossover rather than a sharp transition, consistent with conventional finite-size expectations.
The explained variance spectra in panels (a-ii) and (b-ii) show that the first few principal components capture the majority of the variance, while higher components contribute only marginally. This demonstrates that the complex correlation matrices are effectively governed by a small number of dominant collective modes. These modes are precisely those that dominate the conventional charge structure factor at small and antiferromagnetic wavevectors, establishing an explicit correspondence between PCA modes and traditional correlation analysis.
Importantly, the relative weight carried by the leading components evolves with $U$: at weak coupling, variance is more evenly distributed, reflecting itinerant charge fluctuations, whereas at strong coupling the first component dominates, indicating suppression of charge motion and the emergence of localized charge patterns characteristic of the Mott regime. The fact that this behavior appears similarly in both the $3\times 3$ and $2\times 4$ clusters demonstrates that the PCA fingerprints of the crossover are robust against geometry and spin-sector differences.

\begin{figure}[h]
\centering
\includegraphics[scale=0.42]{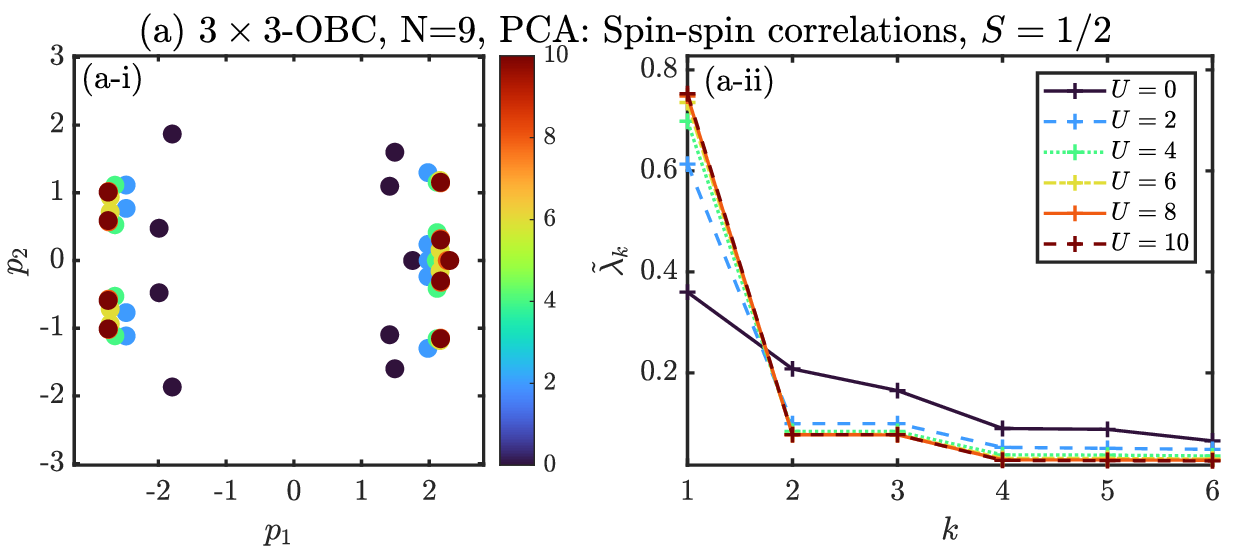}
\quad
\includegraphics[scale=0.42]{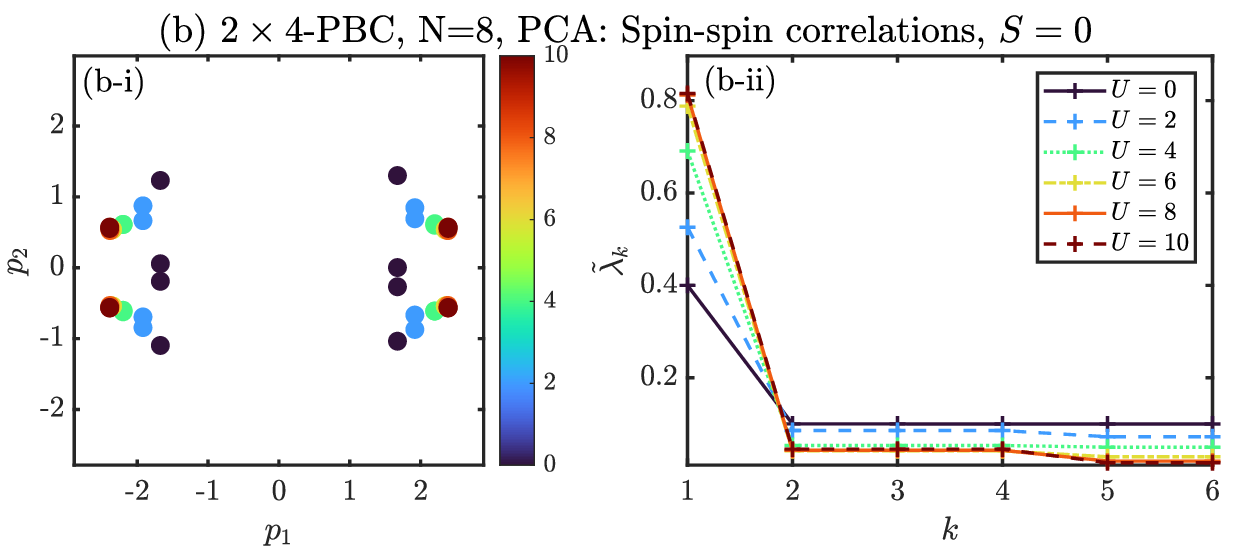}
\caption{Principal component analysis of spin-spin correlation matrices. (a) $3\times 3$ cluster with OBC at half-filling in the $S=1/2$ sector. Panel (a-i) shows the projections of correlation data onto the first two principal components $p_1$ and $p_2$, with color indicating interaction strength $U$. Panel (a-ii) displays the explained variance ratios $\tilde{\lambda}_k$ of the leading six components for several values of $U$. (b) $2\times 4$ cluster with cylindrical boundary conditions at half-filling in the $S=0$ sector. Panels (b-i) and (b-ii) are defined analogously.}
\label{fig:spinpca}
\end{figure}

Figure \ref{fig:spinpca} displays the PCA of spin-spin correlation matrices. In contrast to the charge sector, the spin correlations show an even stronger dimensional reduction: at intermediate and strong coupling, the first principal component alone captures the majority of the variance. This reflects the rapid emergence of a dominant antiferromagnetic fluctuation mode as local moments form.
In the $p_1$-$p_2$ plane, data points at different $U$ values, cluster tightly along a single direction, indicating that once local moments are established, further increases in $U$ only rescale the amplitude of essentially the same spin-correlation pattern. The explained variance ratios confirm this: the leading PCA mode grows sharply beyond intermediate $U$, while all higher components become negligible. 

The same behavior is observed in both cluster geometries, demonstrating that the identification of the dominant antiferromagnetic fluctuation mode is robust against finite-size details. This establishes PCA as a powerful complement to traditional structure-factor analysis: while the latter identifies ordering wavevectors, PCA additionally quantifies the relative statistical weight of competing fluctuation channels and their evolution with interaction strength.

\begin{figure*}[!htbp]
\centering
\includegraphics[scale=0.45]{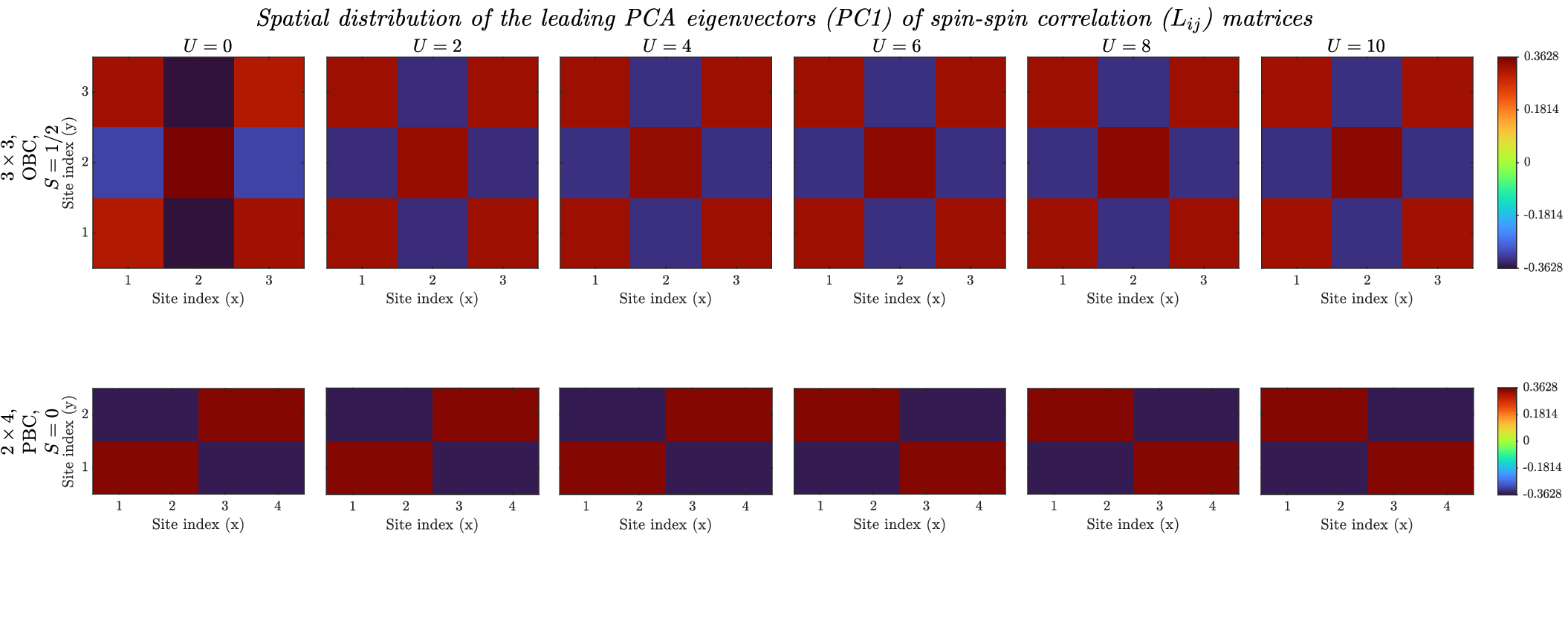}
\caption{Spatial distribution of the leading principal component (PC1) eigenvector of the spin-spin correlation matrix $L_{ij}$. Top row: $3\times 3$ cluster with OBC at half-filling in the $S=1/2$ sector. Bottom row: $2\times 4$ cluster with PBC at half-filling in the $S=0$ sector. Each panel corresponds to a different interaction strength $U$. Color scale indicates the amplitude and sign of the PC1 eigenvector components on each lattice site.}
\label{fig:eigvec}
\end{figure*}

To directly connect the data-driven PCA modes to physical correlation patterns, we visualize in Fig. \ref{fig:eigvec} the real-space structure of the leading PCA eigenvector (PC1) of the spin-spin correlation matrix for representative half-filled clusters. At weak coupling the PC1 eigenvector exhibits a staggered sign structure, which continuously evolves into a robust checkerboard pattern at large $U$. This pattern corresponds precisely to antiferromagnetic correlations at wavevector $(\pi,\pi)$, i.e., the dominant magnetic fluctuation expected in the half-filled Hubbard model. 
With increasing interaction strength, the spatial structure of PC1 remains essentially unchanged while its explained variance grows rapidly. This demonstrates that the Mott crossover is not accompanied by a qualitative change in the dominant spin-correlation pattern, but rather by a strong enhancement of its statistical weight. In other words, PCA identifies the emergence and amplification of antiferromagnetic fluctuations in a fully unsupervised manner.
Notably, the same PC1 spatial pattern is obtained in both the $3\times 3$ OBC and $2\times 4$ PBC clusters, confirming that the extracted dominant fluctuation mode is robust against finite-size geometry and boundary conditions. This directly addresses concerns that PCA results might reflect arbitrary mathematical decompositions: the leading PCA eigenvector reproduces the physically expected antiferromagnetic structure factor pattern, thereby establishing a transparent correspondence between PCA modes and conventional correlation diagnostics. 
This real-space visualization thus provides a clear physical interpretation of the PCA results and highlights the advantage of PCA in automatically extracting ordering tendencies directly from raw correlation data without imposing prior assumptions.

\subsection{Quantum geometry distance metric}
Having established that the leading PCA modes of the correlation matrices encode physically transparent fluctuation patterns, we next turn to a complementary perspective based on quantum geometry. While PCA identifies dominant collective modes and their evolution with interaction strength, it does not directly quantify how rapidly the many-body ground state reorganizes as system parameters are varied. To capture this aspect, we analyze distance measures between ground states at neighboring interaction strengths, which provide a metric-based view of correlation-driven state-space restructuring. This quantum geometric approach thus offers an independent and sensitive diagnostic of the metallic-to-Mott crossover, allowing us to pinpoint regions where the ground-state wave function undergoes its most pronounced reconfiguration, and to correlate these with the emergence of dominant fluctuation modes revealed by the PCA analysis.

To complement our order-parameter and PCA analyses with a wavefunction-based diagnostic, we employ the quantum geometry distance matrix introduced by Hassan {\textit{et. al.} \cite{Hassan2018}. For a normalized many-body ground state $|\psi_0\rangle$, we define a fermionic swap operator $\hat{E}_{ij}$ that exchanges the occupations of single-particle modes $i$ and $j$, satisfying $\hat{E}_{ij}c_{i\sigma}\hat{E}_{ij}^{-1}=c_{j\sigma}$. The associated quantum overlap $\langle \psi_0|\hat{E}_{ij}|\psi_0\rangle$ measures the indistinguishability of the two modes in the correlated ground state. The corresponding distance
\begin{equation}
d_{ij}=\sqrt{1-|\langle \psi_0|\hat{E}_{ij}|\psi_0\rangle|^2},
\end{equation}
defines a proper metric on the space of single-particle modes. For the spinful Hubbard model, we use the swap operator that simultaneously exchanges both spin projections on sites $i$ and $j$, yielding a distance matrix sensitive to combined charge-spin coherence. In this framework, $d_{ij}=0$ indicates complete delocalization (itinerant regime), while $d_{ij}\rightarrow 1$ signals maximal mode distinguishability, characteristic of the localized Mott regime.

\begin{figure}[h]
\centering
\includegraphics[scale=0.35]{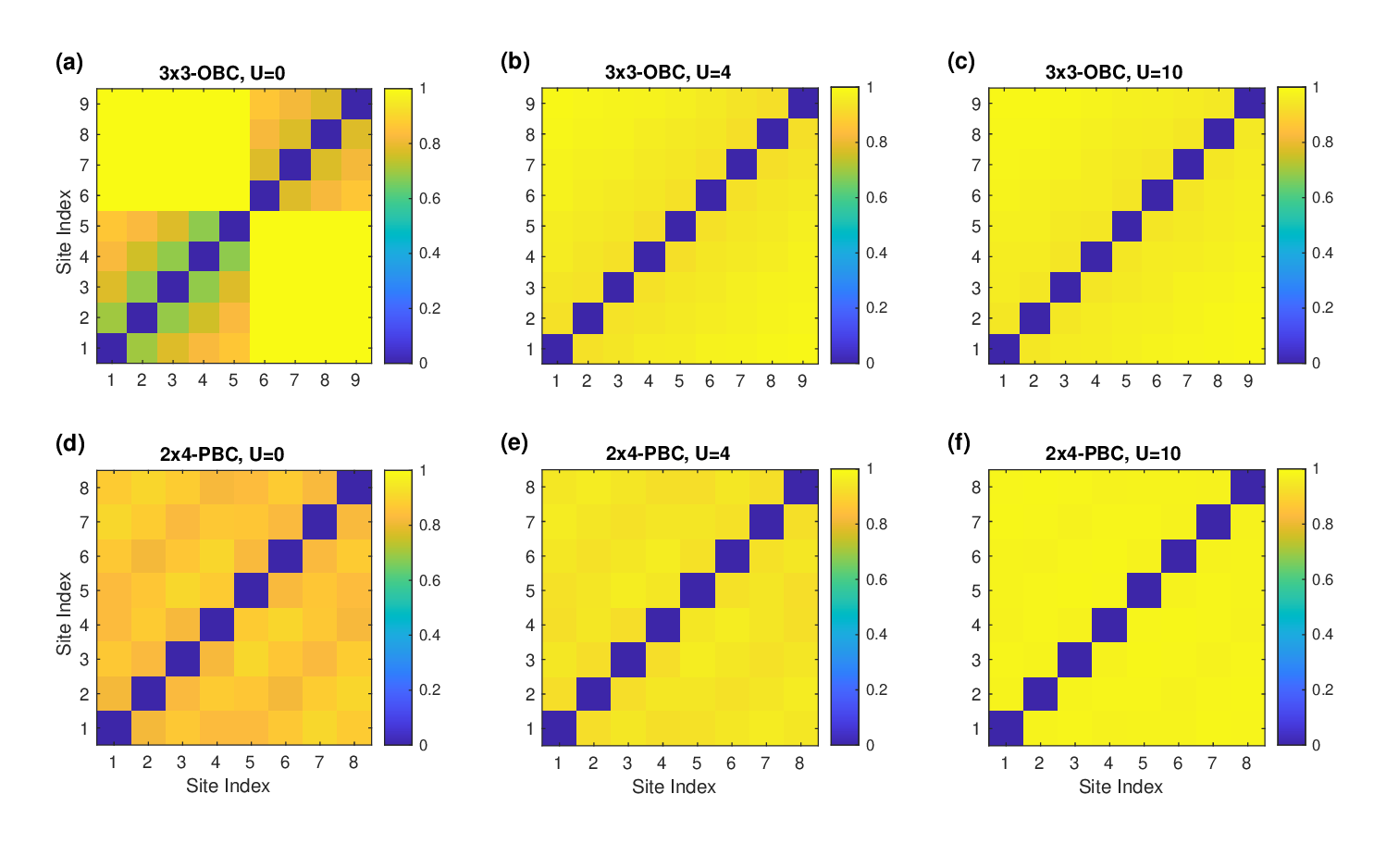}
\caption{Quantum geometry distance matrices $d_{ij}$ for half-filled Hubbard clusters. (a-c) $3\times 3$ OBC cluster in the $S=1/2$ sector for $U=0,4,10$. (d-f) $2\times 4$ PBC cluster in the $S=0$ sector for the same interaction strengths. The evolution from broadly distributed small distances at $U=0$ to sharply diagonal, near-maximal distances at large $U$ signals the crossover from an itinerant metallic regime to a localized Mott-insulating regime.}
\label{fig:distance}
\end{figure}

We now turn to a fully wavefunction-based characterization of correlation-driven localization using the quantum geometry distance matrix. Figures \ref{fig:distance}(a-f) show the distance matrices $d_{ij}$ for the half-filled $3\times 3$ OBC cluster in the $S=1/2$ sector and the $2\times 4$ PBC cluster in the $S=0$ sector.
In the noninteracting limit $U=0$, both clusters exhibit broadly distributed, relatively small distances between many site pairs, reflecting the high degree of mode indistinguishability expected in an itinerant metallic state where electrons are delocalized across the cluster. As interactions increase to intermediate $U=4$, off-diagonal distances grow and the matrix progressively develops a dominant diagonal structure, indicating suppression of intersite coherence and the onset of local moment formation. In the strong-coupling regime $U=10$, the distance matrices become nearly diagonal with $d_{ij}\approx 1$ for $i\neq j$, demonstrating that different sites become maximally distinguishable in the many-body wavefunction --- a direct hallmark of Mott localization.

Further we observe that, despite differences in geometry and boundary conditions, both clusters display the same qualitative evolution of the distance matrix, confirming that the quantum geometry metric provides a robust, basis-independent visualization of the interaction-driven metal-insulator crossover in finite Hubbard clusters.

\section{Summary and Conclusion}
\label{summary}
In summary, we have performed an exact diagonalization study of finite Hubbard clusters to elucidate how electronic correlations reorganize across the interaction-driven metallic to Mott-insulating crossover. By combining conventional observables which includes spin-resolved charge gaps, local moments, double occupancy and correlation functions with modern diagnostics such as entanglement entropy, principal component analysis and quantum-geometry-based distance metrics, we obtained a coherent and multi-perspective characterization of correlation effects in finite systems.

At half-filling, we observed a monotonic growth in the spin-resolved charge-gaps with interaction strength $U$, accompanied by a clear spin-dependent splitting of charge excitation channels. In the strong-coupling regime, the difference between the charge-gaps corresponding to the lowest-spin and the next higher-spin channels follows the effective exchange energy scale $J\sim 4t^2/U$, demonstrating that residual spin dynamics govern the spin structure of charge excitations even when charge fluctuations are suppressed. Comparison between clusters with singlet-spin and finite-spin ground-state reveal that enhanced singlet correlations stabilizes the charge-gap corresponding to the lowest-spin channel. Upon one-hole doping, the collapse of the spin-resolved charge-gaps and persistence of low-energy particle-number fluctuations signal the emergence of compressible, correlated metallic behavior.

Correlation functions and structure factors confirm the growth of the antiferromagnetic spin correlations and suppression of charge coherence with increasing interaction strength. Entanglement measures reveal the redistribution of quantum correlations across interaction regimes, while PCA extracts dominant collective fluctuation modes from full correlation matrices in an unbiased manner. Finally, quantum-geometry-based distance matrices provide a wavefunction-based metric signature of the crossover from delocalized metallic state to localized Mott insulating state.

Altogether, our study establishes a unified and internally consistent microscopic picture of spin-charge interplay, effective exchange-controlled excitation structure and correlation-driven crossover phenomena in finite Hubbard clusters. Our results show that even small clusters capture the essential characteristics of Mott physics and provide a controlled setting for developing and assessing data-driven and quantum-information-based diagnostics of correlated electron systems.

\bibliography{manuscript}
\end{document}